\documentclass[reprint,amsmath,amssymb,aps,pra,superscriptaddress,nofootinbib,longbibliography]{revtex4-1}
\usepackage{graphicx}
\usepackage{xcolor}
\usepackage{bm}
\usepackage{mathrsfs} 
\usepackage{dsfont}

\usepackage[colorlinks=true,breaklinks=true,allcolors=blue]{hyperref}
\usepackage{url}

\newcommand\one{{\mathds{1}}}

\newcommand{\A}{\mathscr{A}}
\newcommand{\C}{\mathscr{C}}

\newcommand{\Com}[2]{\left[{#1},{#2}\right]}

\newcommand{\average}[1]{\left<{#1}\right>}
\renewcommand{\L}{\mathcal{L}}
\newcommand{\mvec}[1]{\boldsymbol #1}
\DeclareMathOperator{\Tr}{Tr}

\newcommand\CC{{\mathds{C}}}

\begin{document}
\title{Quantum effects in the cooperative scattering of light by atomic clouds}

\author{Lorenzo Pucci} 
\affiliation{National Institute for Theoretical Physics (NITheP), Stellenbosch 7600, South Africa} 

\author{Analabha Roy} 
\affiliation{National Institute for Theoretical Physics (NITheP), Stellenbosch 7600, South Africa} 

\author{Tiago Santiago \surname{do Espirito Santo}}
\affiliation{Instituto de F\'{\i}sica de S\~ao Carlos, Universidade de S\~ao Paulo, 13560-970 S\~ao Carlos, SP, Brazil}

\author{Robin Kaiser}
\affiliation{Universit\'e de Nice Sophia Antipolis, CNRS, Institut de Physique de Nice, UMR 7335, F-06560 Valbonne, France}

\author{Michael Kastner} 
\affiliation{National Institute for Theoretical Physics (NITheP), Stellenbosch 7600, South Africa} 
\affiliation{Institute of Theoretical Physics, Department of Physics, University of Stellenbosch, Stellenbosch 7600, South Africa}

\author{Romain Bachelard} 
\email{bachelard.romain@gmail.com} 
\affiliation{Instituto de F\'{\i}sica de S\~ao Carlos, Universidade de S\~ao Paulo, 13560-970 S\~ao Carlos, SP, Brazil}

\date{\today}
\begin{abstract}
Scattering of classical light by atomic clouds induces photon-mediated effective long-range interactions between the atoms and leads to cooperative effects even at low atomic densities. We introduce a novel simulation technique that allows us to investigate the quantum regime of the dynamics of large clouds of atoms. We show that the fluorescence spectrum of the cloud can be used to probe genuine quantum cooperative effects. Signatures of these effects are the occurrence, and the scaling behavior, of additional sidebands at twice the frequency of the classical Mollow sidebands, as well as an asymmetry of the Mollow triplet. 
\end{abstract}

\maketitle

\section{Introduction}

Our understanding of the quantum dynamics of many-body systems has benefited from a number of recent achievements. On the experimental side, cold atom systems and ion traps have reached an unprecedented level of control and allow for the emulation of a large variety of many-body Hamiltonians of interest, including the possibility of tuning coupling parameters \cite{Simon_etal11,Britton2012,Islam_etal13,GeorgescuAshhabNori14}. On the numerical side, progress in the understanding of matrix product state has boosted density matrix renormalization group and other methods, applicable primarily to one-dimensional quantum systems. Reliable simulations of higher-dimensional systems are in general more difficult, and in many cases impossible.

A three-dimensional setting where many exciting discoveries have been made, but also many open questions remain, is light scattering by large clouds of atoms. The classical, linear optics regime of such systems has been extensively studied, and many-body effects, such as superradiance~\cite{Dicke1954,Oliveira2014},  modification of the radiation pressure force \cite{Courteille2010,Bender2010} or cooperative frequency shifts~\cite{Friedberg1973,Rohlsberger2010,Keaveney2012,Jenkins2016,Bromley2016}, were reported. The effective coupling between the atoms in the cloud, mediated by the photon field, turns out to be long-ranged, and as a result cooperative effects occur even in dilute clouds, including superradiance~\cite{Roof2016,Araujo2016}, Dicke subradiance~\cite{Guerin2015}, and spectral broadening~\cite{Sutherland2015}. The strength of the cooperative effects depends in these cases on the cloud optical thickness, not the spatial density.

Beyond the linear optics regime, saturation effects give the atoms a nonlinear response to the electric field of the light. When many atoms are simultaneously in the excited state, the classical coupling of these nonlinear oscillators leads, for example, to the modification of the line shape of the atomic resonance~\cite{Zhu2016}. Entering the field of quantum optics, two-time correlations of the radiated field provide information on the optical coherence of the first kind $g^{(1)}$ and on the fluorescence spectrum. For a single atom, saturation leads to the emergence of the Mollow triplet, a trio of spectral lines of inelastically scattered light, one around the laser frequency and two symmetric ones shifted by the generalized Rabi frequency~\cite{Mollow1969,Schuda1974}; see Fig.~\ref{fig:SpecBabies} (left) for an illustration. In a four-wave mixing configuration, these symmetric bands can be amplified as the wave propagates within the cloud~\cite{agarwal1988,traverso2017}. Furthermore, this redistribution of frequencies is iterated as two scatterers interact through their radiation~\cite{Senitzky1978,Agarwal1980}, which, for dense atomic clouds, results in the presence of additional sidebands at twice the Rabi frequency for pairs of atoms closer than a wavelength~\cite{Ben-Aryeh1988}. In the many-body case, quantum correlations become essential when trying to understand the quantum features of the collective response of the system. This quantum regime of cooperative light scattering in free space is essentially unexplored. The main obstacle for a quantum mechanical treatment is, as usual, the exponential growth of the Hilbert space with the number of atoms. Additionally, and different from, e.g., optical cavities, no suitable collective operators are known for describing the dynamics effectively. A recent first study of quantum effects in the light scattering by dilute atomic clouds made use of perturbative techniques, valid in the regime of strong driving, and was able to predict asymmetries in the scattering spectrum, but no additional spectral lines beyond those of a single atom~\cite{Ott2013}.

In this paper we report the discovery of quantum cooperative effects in light scattering by large dilute clouds of atoms. We show theoretically that quantum correlations building up in the cloud induce cooperative sidebands in the many-body fluorescence spectrum at twice the Rabi frequency, and also lead to a cooperative spectral asymmetry in the Mollow triplet. Both effects scale with the optical thickness, which is a hallmark of their cooperative character. Investigating the angular dependence of the scattering spectrum, we find that quantum cooperativity is more easily detected at large scattering angles, and not in the forward direction. These results provide guidance on where to look for quantum cooperative effects in light scattering experiments on atomic clouds, and are expected to be relevant also for neutral-atom optical clocks, Rydberg atoms, and other settings where effective long-range interactions play a role.

This study of nonperturbative quantum effects in fairly large three-dimensional atomic clouds became possible by a novel simulation technique, combining a discrete phase-space representation of spins \cite{Wootters87,Schach} with higher-order semi-classical evolution equations \cite{PucciRoyKastner16} extended to driven-dissipative Lindblad dynamics. The method is highly accurate for higher-dimensional systems with long-range interactions, and therefore a perfect fit for the problem at hand.
  
\section{Modeling the atomic cloud}

For our purposes the cloud of atoms is modelled as an assembly of $N$ two-level systems at fixed, but usually random, positions $\mathbf{r}_i$ in three-dimensional space. Using a scalar light model, which is valid for dilute clouds, each two-level system can be described by Pauli spin operators $\sigma^\pm=(\sigma^x\pm i\sigma^y)/2$ and $\sigma^z$. This model provides a good description of dilute clouds of atoms cooled below the Doppler limit and trapped in a magneto-optical or dipolar trap. Transitions between the two levels of each atom are driven by a classical planar-wave laser light-field of wave vector $\mathbf{k}_0$, Rabi frequency $\Omega_0$, and detuning $\Delta_0=\omega_0-\omega_a$ from the optical transition. The dynamics is then described by a set of equations that couple the spin degrees of freedom to the photon field \cite{Lehmberg1970}. Performing the rotating-wave, Born and Markov approximations and eliminating the photon degrees of freedom,  equations of motion can be derived for the atomic internal degrees of freedom in the Heisenberg picture~\cite{Lehmberg1970,Ott2013},
\begin{subequations}
\begin{align}
  \frac{d\sigma_{j}^-}{dt} =&
  \left(i\Delta_0-\frac{\Gamma}{2}\right)\sigma_j^- +\frac{i\Omega_0}{2}e^{i \mathbf{k}_0\cdot
  \mathbf{r}_j}\sigma_{j}^z\nonumber
  \\ &+\frac{\Gamma}{2}
  \sum_{m\neq j}^N\sigma_{j}^z\sigma_m^-(\gamma_{jm}-i\Delta_{jm}),\label{s1:M2}\\
  \frac{d\sigma_{j}^z}{dt} =&  i\Omega_0\left(e^{-i \mathbf{k}_0\cdot
  \mathbf{r}_j}\sigma_{j}^- -\text{h.c.}\right)
  -\Gamma(1+\sigma_j^z)\nonumber
  \\&-\Gamma\sum_{m\neq j}^N(\sigma_{m}^+ \sigma_j^-(\gamma_{jm}+i\Delta_{jm})+ \text{h.c.}),
  \label{s3:M2}
\end{align}
\end{subequations}
where the coefficients 
\begin{equation}\label{e:coupling}
\gamma_{jm}=\frac{\sin(k_0|\mathbf{r}_j-\mathbf{r}_m|)}{k_0|\mathbf{r}_j-\mathbf{r}_m|},\quad\Delta_{jm}=\frac{\cos(k_0|\mathbf{r}_j-\mathbf{r}_m|)}{k_0|\mathbf{r}_j-\mathbf{r}_m|}
\end{equation}
describe the spatial dependence of the light-mediated long-range coupling between the atoms, and $\Gamma$ is the transition linewidth of the two-level atoms. Note that the dilute regime allows to neglect near-field terms, which scale as $1/r^3$, and thus focus on long-range effects.

The quantity we want to study, and which is accessible in light-scattering experiments, is the fluorescence spectrum
\begin{equation}\label{S}
    S(\omega)=\lim_{T\rightarrow\infty}\lim_{t\rightarrow\infty}
    \int_{-T}^T d\tau g^{(1)}(t,\tau)e^{-i\omega\tau},
\end{equation}
defined as the Fourier transform of the first-order optical coherence
\begin{equation}\label{eq:g1}
g^{(1)}(t,\tau)=\frac{\langle E(\hat{\mvec{n}},t)E^\dagger(\hat{\mvec{n}},t+\tau)\rangle}
    {\langle E(\hat{\mvec{n}},t)E^\dagger(\hat{\mvec{n}},t)\rangle}
\end{equation}
in the steady state [realized by the limit $t\to\infty$ in \eqref{S}]. The optical coherence is defined in terms of the electric field operator $E$ which, in the far-field limit and emitted in the direction of the normalized vector $\hat{\mvec{n}}$, is given by
\begin{equation}\label{eq:E}
E^\dagger(\hat{\mvec{n}},t)\propto\sum_{j=1}^N\sigma_j^-(t)e^{-ik\hat{\mvec{n}}\cdot \mathbf{r}_j}.
\end{equation}
From Eqs.~\eqref{S}--\eqref{eq:E} one can read off that the crucial nontrivial ingredient for calculating the fluorescence spectrum are the two-time spin--spin correlation functions $\langle\sigma_j^+(t)\sigma_j^-(t+\tau)\rangle$, time-evolved under the evolution equations \eqref{s1:M2} and \eqref{s3:M2}.

\section{Numerical method}

An analytic calculation of the fluorescence spectrum $S$ was reported in Refs.~\cite{Senitzky1978,Agarwal1980} for the case of two atoms pumped at resonance $\omega_a$. In that work, sidebands at frequencies $\omega_a\pm 2\Omega$ were shown to rise in the spectrum, yet the effect is substantial only for atoms closer than a wavelength. A similar effect was predicted for pairs of quantum dots~\cite{Angelatos2015}. Treating more than two atoms is much harder. An exact numerical evaluation of the evolution equations \eqref{s1:M2} and \eqref{s3:M2} is possible for about a dozen atoms or so, but these numbers are too small to investigate scaling behavior or extrapolate to experimentally relevant system sizes.

Here we deal with these difficulties by developing a simulation method for driven-dissipative quantum mechanical evolution equations. Our method builds upon a simulation technique that makes use of a discrete phase-space representation of the Pauli operators \cite{Wootters87,Schach}, and time-evolves phase space points as well as correlation coefficients according to semiclassical evolution equations \cite{PucciRoyKastner16}. We developed these techniques further by extending them to nonunitarily-evolving driven-dissipative systems, and made the method applicable to the computation of two-time correlation functions by making use of the quantum regression theorem. The main features of the method are: (i) The use of semiclassical time-evolution equations makes this method particularly suitable for systems with long-range interactions and systems in higher spatial dimension. (ii) The explicit incorporation of correlation coefficients in the quasiclassical equations of motion allows for the accurate computation of correlation functions; for $N=2$ the exact results are recovered. (iii) The numerical cost scales polynomially, not exponentially, with the system size $N$; hundred and more atoms can be treated.

The novel numerical method, described in the following, is developed for calculating the fluorescence spectrum \eqref{S} of a cloud of two-level atoms as described by the time-evolution equations \eqref{s1:M2} and \eqref{s3:M2}, but is also applicable, more generally, for the calculation of two-time correlation functions in long-range spin models. The atomic cloud we consider is three-dimensional, and the atom-–atom interactions are long-ranged, which makes the system particularly suitable for simulation methods based on the quasiclassical time-evolution of discrete phase-space points~\cite{Schach,PucciRoyKastner16}. However, these methods are formulated for unitarily evolving quantum spin systems and give access to equal-time correlation functions, but not to two-time correlations. In Sec.~\ref{ssec:QuantumRegression} we rewrite the quantities needed for calculating the fluorescence spectrum in a form that is more amenable to a semi-classical time-evolution. In Sec.~\ref{ssec:Time-evolution_equations} we derive semi-classical equations of motion for driven-dissipative, non-unitarily evolving quantum spin systems. In Sec.~\ref{ssec:calculation_V} we introduce a discrete phase space representation and apply it to the calculation of the two-time correlation functions that are required for obtaining the fluorescence spectrum. The resulting novel simulation scheme is benchmarked against exact results in Appendix~\ref{appx:benchmarking}.

\subsection{Quantum regression}
\label{ssec:QuantumRegression}

We extract the spectral properties of the radiated light in direction $\hat{\mvec{n}}$ from the first order optical coherence $g^{(1)}(t,\tau)$ \eqref{eq:g1} by evaluating the expression for the spectrum \eqref{S}, which can be rewritten as
\begin{multline}\label{spectrum}
S(\omega)=\lim_{T\rightarrow\infty}\lim_{t\rightarrow+\infty}\int_{-T}^Td\tau e^{- i\omega\tau}\\
\times\left[g^{(1)}(t,\tau)\Theta(\tau)+g^{(1)}(-t,\tau)\Theta(-\tau)\right],
\end{multline}
where $\Theta$ denotes the Heaviside step function. $g(-t,\tau)$ is a short-hand notation for flipping the sign of $t$ of the unitary terms in \eqref{s3:M2} while keeping the dissipative ones unchanged, which corresponds to time-reversing the dynamics of the total system (two-level atoms plus photons) before eliminating the photonic field. Taking the limit $t\rightarrow+\infty$ inside the integral, we see that $\lim_{t\to\infty}g^{(1)}(t,\tau)$ and $\lim_{t\to\infty}g^{(1)}(-t,\tau)$ are the two limits required for calculating $S$. Here, for being definite, we discuss the first case only, as the second can be treated analogously. According to Eqs.~\eqref{eq:g1} and \eqref{eq:E}, the optical coherence $g^{(1)}(t,\tau)$ can be expressed in terms of two-time correlations of Pauli spin operators, and hence our main object of interest will be $\lim_{t\to\infty}\average{\sigma_ i^a(t)\sigma_ j^b(t+\tau)}$ for all $a,b\in\{x,y,z\}$.

Making use of the quantum regression theorem \cite{Gardiner,Steck}, we can write the two-time spin--spin correlation as
\begin{equation}\label{evol_2r_corr}
\average{\sigma_ i^a(t)\sigma_ j^b(t+\tau)}=\Tr\left\{V(t+\tau,t)\left[\left(V(t,0)\rho\right)\sigma_ i^a\right]\sigma_ j^b\right\},
\end{equation}
where $\rho$ is the initial state and $V(t,t_0)$ denotes the time-evolution operator corresponding to the equations of motion \eqref{s1:M2} and \eqref{s3:M2}. The coefficients of these differential equations do not explicitly depend on time, which implies $V(t+t_0,t_0)=V(t,0)$ for all $t$ and $t_0$. Under this evolution and in the limit $t\to\infty$, we expect $\rho$ to evolve towards a steady state, which we denote by
\begin{equation}
\rho_\text{ss}=\lim_{t\to\infty}V(t,0)\rho.
\end{equation}
This allows us to write the long-time limit of the two-time correlation function as
\begin{equation}\label{limcorr}
\lim_{t\to\infty}\average{\sigma_ i^a(t)\sigma_ j^b(t+\tau)}=\Tr\left[\sigma_ j^b V(\tau,0)\left(\rho_\text{ss}\sigma_ i^a\right)\right].
\end{equation}
The nontrivial object in this expression is the time-evolved operator $V(\tau,0)\left(\rho_\text{ss}\sigma_ i^a\right)$, which we are going to calculate in Sec. \ref{ssec:calculation_V}.

\subsection{Time-evolution equations}
\label{ssec:Time-evolution_equations}

As a starting point for calculating $V(\tau,0)\left(\rho_\text{ss}\sigma_ i^a\right)$ we need the steady state density operator $\rho_\text{ss}$, which we calculate approximately by a method described in the following. This method is applicable to arbitrary $N$-spin trace-1 operators $\A_{1\ldots N}$ under the dynamics generated by a Lindblad operator. Starting from the initial density operator, $\A_{1\ldots N}=\rho$, will allow us to obtain $\rho_\text{ss}$ after sufficiently long evolution times. Other choices of $\A_{1\ldots N}$ will be used to calculate the time-evolution of other trace-1 operators that appear when calculating $V(\tau,0)\left(\rho_\text{ss}\sigma_ i^a\right)$ in Sec. \ref{ssec:calculation_V}.

The propagator $V$ that induces the evolution equations \eqref{s1:M2} and \eqref{s3:M2} can be written as a Lindblad differential equation
\begin{equation}\label{VNeqdWA}
i\partial_t\A_{1\ldots N}=\L\A_{1\ldots N},
\end{equation}
where the Lindblad operator
\begin{equation}\label{e:Lindblad}
\L=\sum_ i\L_ i+\sum_{ i j}\L_{ i j}
\end{equation}
consists of on-site terms
\begin{multline}
\L_ i\A_{1\ldots N}=-\frac{\Delta_0}{2}\Com{\sigma_ i^z}{\A_{1\ldots N}}\\
+\frac{\Omega_0}{2}\Com{e^{- i\mathbf{k}_0\cdot\mathbf{r}_ i}\sigma_ i^-+e^{ i\mathbf{k}_0\cdot\mathbf{r}_ i}\sigma_ i^+}{\A_{1\ldots N}}
\end{multline}
and pair interactions
\begin{multline}\label{e:Lij}
\L_{ i j}\A_{1\ldots N}=\Delta_{ i j}\Com{\sigma_ i^+\sigma_ j^-}{\A_{1\ldots N}}+\\
i\gamma_{ij}\left(\sigma_ j^-\A_{1\ldots N}\sigma_ i^+-\tfrac{1}{2}\sigma_ i^+\sigma_ j^-\A_{1\ldots N}-\tfrac{1}{2}\A_{1\ldots N}\sigma_ i^+\sigma_ j^-\right),
\end{multline}
with
\begin{align}
\Delta_{ i j}&=-\frac{\Gamma}{2}\begin{cases}\displaystyle
\frac{\cos{\left(k_0|\mathbf{r}_ i-\mathbf{r}_ j|\right)}}{k_0|\mathbf{r}_ i-\mathbf{r}_ j|} & \text{for $i\neq j$},\\
0 & \text{for $i=j$},
\end{cases}\\
\gamma_{ij}&=\Gamma\begin{cases}\displaystyle
\frac{\sin{\left(k_0|\mathbf{r}_ i-\mathbf{r}_ j|\right)}}{k_0|\mathbf{r}_ i-\mathbf{r}_ j|} & \text{for $i\neq j$},\\
1 & \text{for $i=j$},
\end{cases}
\end{align}
and $\Gamma=d^2k_0^3/(2\pi\hbar\epsilon_0)$. 

Taking partial traces on both sides of \eqref{e:Lindblad} one obtains, in the spirit of the Bogoliubov-Born-Green-Kirkwood-Yvon (BBGKY) hierarchy \cite{Bonitz}, a set of coupled evolution equations,  
\begin{subequations}
	\begin{align}
	i\partial_t \A_i&=\L_i\A_i+\sum_{m\neq i}\Tr_m\L_{im}\A_{im},\label{e:1st_order_app1}\\
	i\partial_t \A_{ij}&=\left(\L_i+\L_j+\L_{ij}\right)\A_{ij}\nonumber\\
	&\quad+\sum_{m\neq i,j}\Tr_m\left(\L_{im}+\L_{jm}\right)\A_{ijm},\label{e:2nd_order_app1}
	\end{align}
\end{subequations}
where the reduced operators
\begin{equation}
\A_i=\Tr_{\{k\neq i\}}\A_{1\ldots N},\qquad
\A_{ij}=\Tr_{\{k\neq i,j\}}\A_{1\ldots N},
\end{equation} 
are defined as partial traces over all sites except the indexed ones. It is expected for the dilute regime that the two-atom coupling dominates over three- or more-atom coupling, which makes a cluster expansion suitable for the truncation of the hierarchy of equations. By means of a cluster expansion we separate the reduced $\A$ operators into product and connected parts,
\begin{subequations}
	\begin{align}
	\label{ClExp} \A_{ i j}&=\A_ i \A_ j+\C_{ i j},\\	
	\label{ClExp2}\A_{ i j m}&=\A_ i \A_ j \A_ m+\A_ i \C_{ j m}+\A_ j \C_{ i m}+\A_ m \C_{ i j}+\C_{ i j m},
	\end{align}
\end{subequations}
which implicitly defines the connected operators $\C$. Substituting these definitions into \eqref{e:1st_order_app1} and \eqref{e:2nd_order_app1}, we rewrite the first two equations of the BBGKY hierarchy as
\begin{subequations}
	\begin{align}
	\label{eqlr} i\partial_t \A_ i&=\L_ i \A_ i+\sum_{ m\neq i}\Tr\left[\L^S_{ i m}(\C_{ i m}+\A_ i \A_ m)\right],\\
	\label{eq2lr}  i\partial_t \C_{ i j}&=(\L_ i+\L_ j)\C_{ i j}+\L^S_{ i j}(\C_{ i j}+\A_ i \A_ j)\\
	&\quad\nonumber-\A_ i\Tr_ i\left[\L^S_{ i j}(\C_{ i j}+\A_ i \A_ j)\right]\\
	&\quad\nonumber-\A_ j\Tr_ j\left[\L^S_{ i j}(\C_{ i j}+\A_ i \A_ j)\right]\\
	&\quad\nonumber+\sum_{ m\neq i, j}\Tr_ m\left[\L^S_{ i m}(\A_ i \C_{ j m}+\A_ m \C_{ i j}+\C_{ i j m})\right]\\
	&\quad+\sum_{ m\neq i, j}\Tr_ m\left[\L^S_{ j m}(\A_ j \C_{ i m}+\A_ m \C_{ i j}+\C_{ i j m})\right],\nonumber
	\end{align}
\end{subequations}
where $\L^S_{ i j}=\L_{ i j}+\L_{ j i}$. Equation~\eqref{eq2lr} contains three-spin connected contributions $\C_{ i j m}$, the time-evolution of which depends on four-spin terms, and so on. To turn this into a numerically tractable problem, we truncate the BBGKY hierarchy at second order by neglecting the terms $\C_{ i j m}$ in \eqref{eq2lr}. As stated in \cite{PucciRoyKastner16}, if we also neglected the $\C_{i j}$ terms, we would recover the classical time-evolution equation presented in \cite{Schach}, in that sense, more terms of the truncated hierarchy means more quantum corrections for the spins dynamics. The two-spin connected contribution $\C_{i j}$  are related to the two-spin quantum correlations, which implies that, unlike in the classical case, the two-spin connected correlations do not vanish, $\langle\sigma_j^{\pm,z}\sigma_m^{\pm,z}\rangle - \langle\sigma_j^{\pm,z}\rangle\langle\sigma_m^{\pm,z}\rangle \neq 0$. This is the main approximation made in the numerical scheme, and it gives good results whenever genuine three- and more-spin connected contributions are negligible.

To bring the resulting truncated operator equations into a numerically tractable form, we expand all operators in terms of Pauli spin operators and we obtain a set of coupled ordinary differential equations (see Appendix~\ref{appx:evolution_pauli_coefficients}), which can be integrated by standard numerical methods.


\subsection{Calculation of \texorpdfstring{$V(\tau,0)\left(\rho_\text{ss}\sigma_i^a\right)$}{}}
\label{ssec:calculation_V}

We calculate the steady-state density operator $\rho_\text{ss}$ by setting to zero the left-hand sides of \eqref{ca_eq1} and \eqref{ca_eq2}, and numerically solving the resulting algebraic equations by a standard Newton-Krylov solver. The stationary values of the a- and c-coefficients in those equations encode the required information on $\rho_\text{ss}$.

Starting from the thus obtained steady-state density operator $\rho_\text{ss}$, we expand $\rho_\text{ss}\sigma_ i^a$ in terms of so-called phase point operators, rewrite the result in terms of trace-1 operators, and then use again the time-evolution equations of Secs.~B and C in order to obtain $V(\tau,0)\left(\rho_\text{ss}\sigma_ i^a\right)$.

The discrete phase-space representation of a single spin-$1/2$ degree of freedom as introduced by Wootters \cite{Wootters87} is based on a discrete phase space
\begin{equation}
\Gamma=\left\{(0,0),(0,1),(1,0),(1,1)\right\}
\end{equation}
consisting of four points, each of which has an associated three-vector, $\mvec{r}_{(0,0)}=(1,1,1)$, $\mvec{r}_{(0,1)}=(-1,-1,1)$, $\mvec{r}_{(1,0)}=(1,-1,-1)$, and $\mvec{r}_{(1,1)}=(-1,1,-1)$. To each phase space point $\alpha\in\Gamma$ one assigns a so-called phase point operator
\begin{equation}\label{e:Aalpha}
A_\alpha = \tfrac{1}{2}(\one+\mvec{r}_\alpha\cdot\mvec{\sigma}),
\end{equation}
where $\mvec{\sigma}=\left(\sigma^x,\sigma^y,\sigma^z\right)$ is the vector of Pauli operators. The phase point operators form a basis, and any operator on $\CC^2$ can be expressed as a linear combination of the four operators $A_\alpha$. Similarly one could expand an operator on the tensor product Hilbert space $(\CC^2)^{\otimes N}$ of $N$ spin-$1/2$ degrees of freedom in the corresponding tensor product basis of $A_\alpha$. Here we follow a different approach and do this expansion only for the $i$th factor of the product space, which yields \footnote{It turns out to be advantageous to expand in an overcomplete basis, using the phase point operators corresponding to $\mvec{r}_{(0,0)}^\prime=(1,-1,1)$, $\mvec{r}_{(0,1)}^\prime=(-1,1,1)$, $\mvec{r}_{(1,0)}^\prime=(1,1,-1)$, $\mvec{r}_{(1,1)}^\prime=(-1,-1,-1)$, in addition to those defined above. While such an expansion gives identical results on an exact level, differences arise when approximating the time-evolution. Avoiding the expansion altogether, and writing $\rho_\text{ss}\sigma_i^a$ directly in terms of trace-$1$ operators, is also feasible, but again turned out to perform worse than the scheme described here.
}
\begin{equation}\label{rsexpan}
\rho_\text{ss}\sigma_i^a=\frac{1}{2}\sum_{\alpha_ i}\Tr_i\left[A_{\alpha_i}\rho_\text{ss}\sigma_i^a\right]A_{\alpha_i},
\end{equation}
where $\Tr_i$ denotes a partial trace over the $i$th factor of the tensor product Hilbert space. By elementary spin algebra, the coefficients of this expansion, which are operator-valued and act on $(\CC^2)^{\otimes(N-1)}$, can be written as
\begin{multline}\label{eqicss}
2\Tr_ i\left[A_{\alpha_ i}\rho_\text{ss}\sigma_ i^a\right]=\Tr_ i\Bigl[\sigma_ i^a(\one_ i+\sum_cr_{\alpha_i}^c\sigma_ i^c)\rho_\text{ss}\Bigr]\\
=\Tr_ i\left(\sigma_ i^a\rho_\text{ss}\right)+r_{\alpha_i}^a\Tr_ i\left(\rho_\text{ss}\right)+ i\sum_{cd}\varepsilon^{acd}r_{\alpha_i}^c\Tr_ i\left(\sigma_ i^d\rho_\text{ss}\right).
\end{multline}
Next we rewrite \eqref{eqicss} as a linear combination of trace-$1$ operators, such that the time-evolution scheme of Sec. \ref{ssec:Time-evolution_equations} can be applied to each of those operators. To this purpose it is convenient to (partially) expand operators in the tensor product basis of Pauli spin operators, where we denote the expansion coefficients by
\begin{equation}\label{e:exprhoss}
s_i^a=\Tr(\sigma_i^a \rho_\text{ss}),\qquad
s_{ij}^{ab}=\Tr(\sigma_i^a\sigma_j^b \rho_\text{ss}).
\end{equation}
Starting from \eqref{eqicss} we can write \footnote{We tested other ways of expressing \eqref{eqicss} in terms of trace-1 operators, but for our purposes none of them turned out to be advantageous in terms of accuracy or computational cost.}
\begin{multline}\label{eqicss_2}
\!\!2\Tr_ i\left(A_{\alpha_ i}\rho_\text{ss}\sigma_ i^a\right)=(1+s_ i^a)\tilde{\rho}^a_{\text{ss},\not{\,i}}+ i\sum_{cd}\varepsilon^{acd}r_{\alpha_i}^c(1+s_ i^d)\tilde{\rho}^d_{\text{ss},\not{\,i}}\\
+\Bigl[(r_{\alpha_i}^a-1)- i\sum_{cd}r_{\alpha_i}^c\varepsilon^{acd}\Bigr]\rho_{\text{ss},\not{\,i}},
\end{multline}
where we have defined
\begin{equation}
\rho_{\text{ss},\not{\,i}}=\Tr_ i\rho_\text{ss},\qquad\tilde{\rho}_{\text{ss},\not{\,i}}^a=\frac{\Tr_i\left[(\one+\sigma_i^a)\rho_\text{ss}\right]}{1+s_i^a},
\end{equation}
both of which are trace-$1$ operators on $(\CC^2)^{\otimes(N-1)}$. Those operators can also be expanded in terms of Pauli spin operators, and the corresponding expansion coefficients can be expressed in terms of the coefficients \eqref{e:exprhoss} of $\rho_\text{ss}$,
\begin{subequations}
	\begin{align}
	\tilde{s}_k^{a_k,a}&=\Tr \left(\sigma_k^{a_k}\tilde{\rho}_{ss,\not{\,i}}^a\right)=\frac{s_k^{a_k}+s_{ i k}^{aa_k}}{1+s_i^a},\\
	\tilde{s}_{jk}^{a_ja_k,a}&=\Tr \left(\sigma_j^{a_j}\sigma_k^{a_k}\tilde{\rho}_{ss,\not{\,i}}^a\right)=\frac{s_{jk}^{a_ja_k}+s_{ i jk}^{aa_ja_k}}{1+s_i^a},
	\end{align}
\end{subequations}
and so on. Inserting \eqref{rsexpan} and \eqref{eqicss_2} into \eqref{limcorr} we obtain
\begin{widetext}
	\begin{multline}\label{explicit_correl}
	\lim_{t\rightarrow\infty}\average{\sigma_ i^+(t)\sigma_ j^-(t+\tau)}=\frac{1}{4}\sum_{\alpha_ i}\bigl\{(1+s_ i^x)(1-r_{\alpha_i}^z)(a_{ j;\alpha_i}^{x;x}(\tau)-ia_{ j; \alpha_i}^{y;x}(\tau))+(1+s_ i^y)(1-r_{\alpha_i}^z)(a_{ j;\alpha_i}^{y;y}(\tau)+ i a_{ j;\alpha_i}^{x;y}(\tau))\\
	+(1+s_ i^z)\left[r_{\alpha_i}^xa_{ j;\alpha_i}^{x;z}(\tau)+r_{\alpha_i}^ya_{ j;\alpha_i}^{y;z}(\tau)+ i(r_{\alpha_i}^ya_{ j;\alpha_i}^{x;z}(\tau)-r_{\alpha_i}^xa_{ j;\alpha_i}^{y;z}(\tau))\right]\\
	+(r_{\alpha_i}^z-1)\left[a_{ j;\alpha_i}^x(\tau)+a_{ j;\alpha_i}^y(\tau)+ i(a_{ j;\alpha_i}^x(\tau)-a_{ j;\alpha_i}^y(\tau))\right]\bigr\}
	\end{multline}
\end{widetext}
with
\begin{subequations}
	\begin{align}
	a_{ j;\alpha_i}^{b;a}(\tau)&=\Tr\left[\sigma_ j^bV(\tau,0)(\tilde{\rho}^a_{ss,\not{\,i}}A_{\alpha_ i})\right],\label{e:coeffs1}\\
	a_{ j;\alpha_i}^b(\tau)&=\Tr\left[\sigma_ j^bV(\tau,0)(\rho_{ss,\not{\,i}}A_{\alpha_ i})\right].\label{e:coeffs2}
	\end{align}
\end{subequations}
According to \eqref{e:coeffs1} and \eqref{e:coeffs2}, in order to obtain the desired two-time correlation function \eqref{explicit_correl}, we need to calculate for each $i$ and each phase space operator $A_{\alpha_ i}$ the time-evolution of four operators, namely $\tilde{\rho}^a_{ss,\not{\,i}}A_{\alpha_ i}$ for $a\in\{x,y,z\}$, and $\rho_{ss,\not{\,i}}A_{\alpha_ i}$. We do this by making use of the method developed in Sec. \ref{ssec:Time-evolution_equations}, letting $\A_{1\ldots N}$ in \eqref{VNeqdWA} take the role of each of the four mentioned operators. The computational cost of the method scales like $3N(3N-1)/2$ with the system size $N$. Applying it to the four trace-$1$ operators, the eight phase point operators, and the $N$ lattice sites required in \eqref{explicit_correl}, \eqref{e:coeffs1}, and \eqref{e:coeffs2}, results in an overall computational cost that scales asymptotically like $N^3$.

\section{Results}

\begin{figure}
\centering
\includegraphics[width=1\linewidth]{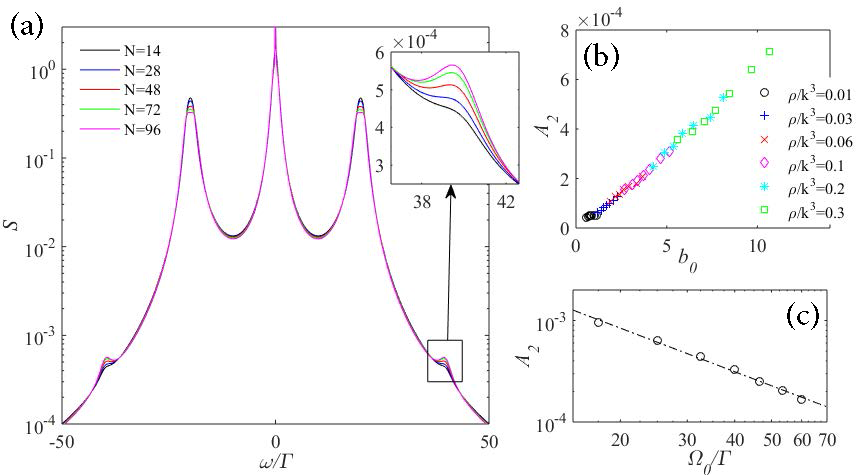}
\caption{\label{fig:SpecBabies} (a) Fluorescence spectrum for a cloud of density $\rho/k^3=0.1$, driven at $\Omega_0=20\Gamma$ at resonance ($\Delta=0$), for $N=14,\ 28,\ 48,\ 72$ and $96$ atoms, with the inset showing the behaviour of the peaks at $\omega\approx2\Omega_0$. Amplitude of the additional Mollow sidebands as a function of (b) the optical thickness $b_0$ for different densities ($\Omega=20\Gamma$, $N=72$ and $\Delta=0$) and (c) of the Rabi frequency ($\rho/k^3=0.1$, $N=72$ and $\Delta=0$). The amplitude is defined as $\int_{2\Omega_0-\delta\omega}^{2\Omega_0+\delta\omega} |S-S_1|d\omega$, where $S_1$ is the single-atom spectrum, and $\delta\omega$ a suitably chosen integration range. The dash-dotted line in (c) refers to a power-law fit ($A_2\approx 0.06(\Omega/\Gamma)^{-1.4}$).}
\end{figure}

We performed simulations for system sizes $N=14$, $28$, $48$, $72$ and $96$ atoms at a fixed low density $\rho$. The atoms are placed at random in a spherical volume, but with the constraint of a minimal distance of one fourth of the mean distance between neighbours, such as to rule out unwanted noncooperative effects due to accidentally close pairs of atoms. Fig.~\ref{fig:SpecBabies}(a) shows the numerically computed fluorescence spectra for various $N$, at fixed density and laser parameters. The three prominent peaks in the plot at $\omega=0$ and $\pm\Omega_0$ form the Mollow triplet~\cite{Mollow1969,Schuda1974}. The first main result is the observation of additional sidebands in the fluorescence spectrum at $\omega=\pm 2\Omega_0$. These sidebands are genuine quantum effects, as they require the presence of quantum pair correlations. Indeed if connected correlations between different sites were absent and two-time correlations would factorize, $\langle\sigma_j^{\pm,z}\sigma_m^{\pm,z}\rangle = \langle\sigma_j^{\pm,z}\rangle\langle\sigma_m^{\pm,z}\rangle$ for $j\neq m$, one would have $\langle\sigma_j^+(t)\sigma_m^-(t+\tau)\rangle = \langle\sigma_j^+(t)\rangle\langle\sigma_m^-(t)\rangle$ in the steady-state regime, and in that case the inelastic ($\omega\neq 0$) spectrum would only depend on single-site two-time correlations $\langle\sigma_j^+(t)\sigma_j^-(t+\tau)\rangle$. The factorizing terms $\langle\sigma_j^{\pm,z}\rangle\langle\sigma_m^{\pm,z}\rangle$ may modify the local Rabi frequency experienced by each atom and inhomogeneously broaden the single-atom Mollow triplet, but cannot give rise to higher-order sidebands. In other words, a classical treatment of the Hamiltonian (or, as in this case, Lindbladian) results in the absence of connected correlations between different spins, and no additional sidebands are observed.

The novel sidebands are true cooperative effects. If the sidebands were two-atom or few-atom effects, their peak height would depend only on the spatial density. In Figs.~\ref{fig:SpecBabies}(a) and (b), however, we observe that the sidebands grow with the number of atoms $N$ even at fixed density, and scale linearly with the optical thickness $b_0=2N/(kR)^2$ rather than with the spatial density.  This effect can be attributed to the long-range nature of the effective interactions \eqref{e:coupling} between the atomic internal degrees of freedom. The scaling with $b_0$ is reminiscent of cooperative phenomena in the linear optics regime~\cite{Guerin2016}, but is here observed, for the first time to the best of our knowledge, for a quantum cooperative phenomenon in free space. Furthermore, although single scattering processes may exhibit quantum optics interferences phenomena~\cite{Grangier1986}, they cannot capture the additional sidebands. These sidebands at $\pm 2\Omega_0$ can be understood as the first step of a higher order harmonic generation process, where the next orders could be studied by including higher-order quantum correlations. However, the low relative intensity of approximately $10^{−3}$ makes it hard to detect the peaks at $\pm 2 \Omega_0$ experimentally. With that in mind, we searched for traces of quantum cooperativity in the Mollow triplet bands ($\pm \Omega_0$).

\begin{figure}
\centering 
\includegraphics[width=0.49\linewidth]{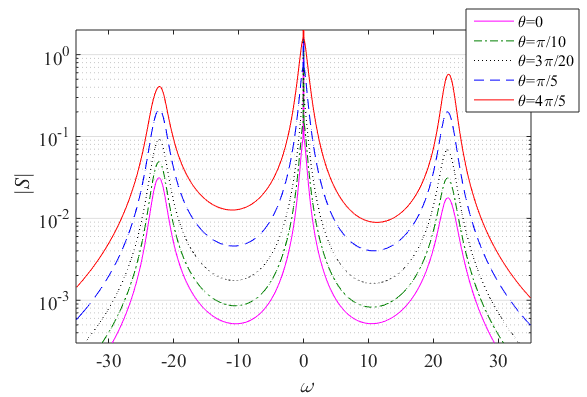}
\includegraphics[width=0.49\linewidth]{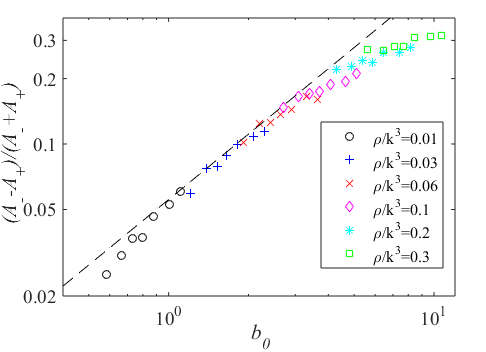}
\caption{\label{fig:Asymmetry} Left: Fluorescence spectrum for an atomic cloud of density $\rho/k^3=0.1$, driven at $\Omega_0=20\Gamma$ out of resonance ($\Delta=-\Omega_0/2$), and for scattering angles $\theta=0$, $\pi/10$, $3\pi/20$, $\pi/5$, $4\pi/5$. The asymmetry of the sidebands is clearly visible. Right: The asymmetry of the spectrum in the forward direction ($\theta = 0$), quantified by $(A_- - A_+)/(A_− +A_+)$ where $A\pm$ is the amplitude of the sideband at $\pm\Omega_0$, plotted as a function of the optical thickness for different densities and system sizes.}
\end{figure}

The second result is the observation that quantum cooperativity breaks the symmetry of the spectrum. The single-atom fluorescence spectrum is always symmetric with respect to the frequency of the driving light, independently of the detuning of the driving from the atomic resonance \cite{Mollow1969}.  For large atomic clouds and in the presence of detuning ($\Delta \neq 0$), it was predicted that coherence effects may induce an asymmetry of the Mollow sidebands in the forward scattering direction \cite{Ott2013}. Our simulations show a similar effect for the scattering of detuned light, where the Mollow sidebands at $\omega=\pm\Omega$ exhibit a significant asymmetry (Fig.~\ref{fig:Asymmetry} left). This asymmetry scales with the optical thickness  $b_0$ (Fig.~\ref{fig:Asymmetry} right), which confirms the cooperative nature of this effect. In the absence of quantum correlations the spectrum, being composed of the sum of $N$ symmetric spectra, is necessarily symmetric, which confirms the genuine quantumness of the observed asymmetry. However, going beyond the prediction of Ref.~\cite{Ott2013}, we here observe that the asymmetry is also present outside the forward lobe, i.e., for scattering angles $\theta\geq1/kR$ where diffuse light dominates (Fig.~\ref{fig:Asymmetry} left). Surprisingly, the asymmetry is inverted in the forward direction ($\theta<1/kR$) in comparison with $\theta>1/kR$. Experimentally the asymmetry of the standard Mollow sidebands, which reaches $\sim30\%$ in our simulations, should be relatively easy to detect.

\begin{figure}
\centering 
\includegraphics[width=0.99\linewidth]{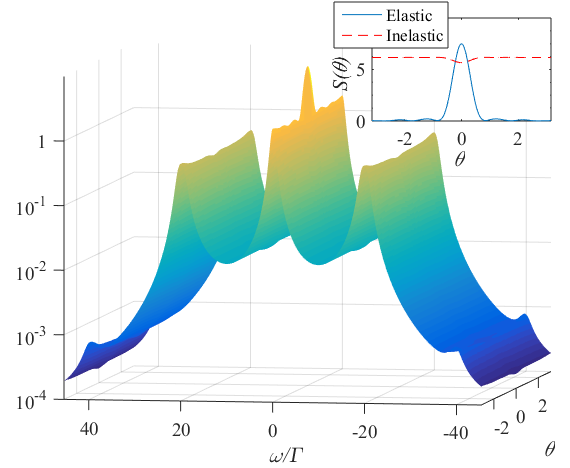}
\caption{\label{fig:Anisotropy} Angular dependence of the fluorescence spectrum of a cloud of $N=72$ atoms with density $\rho=0.1k^3$, driven by a field with $\Omega_0=20\Gamma$ at resonance, which corresponds to a saturation parameter $s=3200$. The inset shows the elastic and integrated inelastic spectra ($S_\text{el}(\theta)=S(\omega=0,\theta)$ and $S_\text{inel}(\theta)=\int_{\omega\neq 0} S(\omega,\theta) d\omega$), respectively, as discussed in the text.} 
\end{figure}

In Fig.~\ref{fig:Anisotropy}  we show the fluoresence spectrum as a function of the scattering angle $\theta$ and the frequency $\omega$ in the regime of deep saturation, where most of the light is expected to be scattered inelastically ($\omega\neq0$). In this regime the portion of elastically scattered light for a single atom goes as $1/s=(\Delta_0^2+\Gamma^2/4)/2\Omega_0^2$ at large $\Omega_0$, $s$ being referred as the saturation parameter, so most of the light is scattered inelastically. We clearly observe the quasi-isotropic inelastic Mollow triplet and higher order sidebands. For the parameters considered, a strong elastic component is particularly visible in the forward direction (see inset), which we attribute to the constructive interference of the elastic component of the electric field in the forward direction (which, according to linear optics, is expected to scale like $N^2$ with the system size). This indicates that signatures of quantum cooperativity, which are intimately connected to inelastic scattering, may be more easily detected at larger scattering angles, and not in the forward direction. We note however that in the forward direction the inelastic component exhibits a small dip. The physical origin of this feature remains to be understood.

\section{Conclusions}

We have reported the discovery of signatures of quantum cooperativity in the fluorescence spectrum of large dilute atomic clouds. The rise of additional sidebands at frequencies $\pm2\Omega_0$ from the central line, as well as the asymmetry in the spectrum of the cloud driven out of resonance, are identified as proper quantum effects that cannot occur in the absence of genuine quantum correlations. Moreover, by analyzing parameter dependences and scaling properties, the cooperative nature of the observed phenomena is revealed. Cooperativity is ultimately related to the long-range character of the effective atom--atom interactions induced by the photon field. The deficiency in inelastically scattered photons in the forward cone ($\lvert\theta\rvert\leq1/kR$) is particularly interesting, because it implies that the forward direction, which has long been considered a natural candidate for probing cooperative phenomena in the linear-optics regime~\cite{Scully2006}, may be less suitable for probing quantum cooperative effects. Furthermore, while the second-order optical coherence is usually considered the ideal candidate for revealing the quantum nature of the light scattering by atoms, with photon bunching~\cite{Morgan1966} and anti-bunching~\cite{Kimble1967} as paradigmatic signatures, we show in this paper that the first-order optical coherence $g^{(1)}$, which witnesses the quantum nature of the atom-light coupling, already shows clear signatures of quantum cooperativity. More generally, our results suggest that the quantum optics regime of an optically deep system is substantially richer than its single-atom physics, and holds much promise for further studies of cooperative effects. This may become relevant for neutral atom optical clocks or other long-range quantum systems such as Rydberg atoms. To gain access to this regime on the computational side, the simulation technique developed in this paper, based on a truncation of the hierarchy of correlations, proves to be a powerful tool.

The cooperative nature of the observed effects suggests that dilute atomic clouds might be used as experimental platforms for quantum-simulating plasmas, free electron lasers, and other quantum long-range interacting systems in which cooperativity plays an essential role.

\begin{acknowledgments}
We thank N. Piovella for stimulating discussions. A.R.\ acknowledges financial support from the National Research Foundation of South Africa.
This work has received financial support from FAPESP and CNPq.
M.K.\ acknowledges financial support from the National Research Foundation of South Africa via the Incentive Funding and the Competitive Programme for Rated Researchers. M.K. and R.B. acknowledge financial support by a FAPESP/SU collaboration grant.
\end{acknowledgments}

\appendix

\section{Time-evolution of the Pauli expansion coefficients}
\label{appx:evolution_pauli_coefficients}

We expand all operators in (\ref{ClExp}) and (\ref{ClExp2}), except $\C_{ i j m}$ that is neglected, in terms of Pauli spin operators,

\begin{equation}\label{Pauli}
\A_{ i}=\frac{1}{2}\left(\one+\mvec{a}_ i\cdot\mvec{\sigma}_ i\right),\qquad\C_{ i j}=\frac{1}{4}\sum_{a,b\in\{x,y,z\}}c_{ i j}^{ab}\sigma_ i^a\sigma_ j^b.
\end{equation}
Inserting these expansions into \eqref{eqlr} and the truncated version of \eqref{eq2lr}, and making use of Lindblad equations \eqref{VNeqdWA}--\eqref{e:Lij}, we obtain time-evolution equations for the Pauli expansion coefficients,
\begin{widetext}
	\begin{multline}\label{ca_eq1}
	\partial_ta_ i^a=\sum_{b}a_ i^b\left\{-\Delta_0\varepsilon^{zba}+\Omega_0\left[\varepsilon^{xba}\cos(-\mathbf{k}_0\cdot\mathbf{r}_ i)+\varepsilon^{yba}\sin(-\mathbf{k}_0\cdot\mathbf{r}_ i)\right]\right\}-\Gamma\left[\tfrac{1}{2}a_i^a(1+\delta^{za})+\delta^{za}\right]\\
	+\sum_{b}\sum_{ m\neq i}\left\{\varepsilon^{xba}\left[\Delta_{ i m}\left(a_ i^ba_ m^x+c_{ i m}^{bx}\right)-\tfrac{1}{2}\gamma_{ i m}\left(a_ i^ba_ m^y+c_{ i m}^{by}\right)\right]+\varepsilon^{yba}\left[\Delta_{ i m}\left(a_ i^ba_ m^y+c_{ i m}^{by}\right)+\tfrac{1}{2}\gamma_{ i m}\left(a_ i^ba_ m^x+c_{ i m}^{bx}\right)\right]\right\}
	\end{multline} 
	and
	\begingroup
	\allowdisplaybreaks
	\begin{align}\label{ca_eq2}
	\partial_tc_{ i j}^{ab}=&\sum_{c}c_{ i j}^{cb}\left\{-\Delta_0\varepsilon^{zca}+\Omega_0\left[\cos(-\mathbf{k}_0\cdot\mathbf{r}_ i)\varepsilon^{xca}+\sin(-\mathbf{k}_0\cdot\mathbf{r}_ i)\varepsilon^{yca}\right]\right\}\\
	&+\sum_{c}c_{ i j}^{ac}\left\{-\Delta_0\varepsilon^{zcb}+\Omega_0\left[\cos(-\mathbf{k}_0\cdot\mathbf{r}_ j)\varepsilon^{xcb}+\sin(-\mathbf{k}_0\cdot\mathbf{r}_ j)\varepsilon^{ycb}\right]\right\}\nonumber\\
	&-\Gamma c_{ i j}^{ab}\left(1+\frac{\delta^{az}+\delta^{bz}}{2}\right)-\gamma_{ij}\sum_{c,d}\left(c_{ i j}^{cd}+a_ i^ca_ j^d\right)\left(\varepsilon^{xca}\varepsilon^{dxb}+\varepsilon^{yca}\varepsilon^{dyb}\right)\nonumber\\
	&+\sum_{c}c_{ i j}^{cb}\sum_{ m\neq i j}\left[a_ m^x\left(\Delta_{ i m}\varepsilon^{xca}+\frac{\gamma_{ i m}}{2}\varepsilon^{yca}\right)+a_ m^y\left(\Delta_{ i m}\varepsilon^{yca}-\frac{\gamma_{ i m}}{2}\varepsilon^{xca}\right)\right]\nonumber\\
	&+\sum_{c}c_{ i j}^{ac}\sum_{ m\neq i j}\left[a_ m^x\left(\Delta_{ j m}\varepsilon^{xcb}+\frac{\gamma_{ j m}}{2}\varepsilon^{ycb}\right)+a_ m^y\left(\Delta_{ j m}\varepsilon^{ycb}-\frac{\gamma_{ j m}}{2}\varepsilon^{xcb}\right)\right]\nonumber\\
	&+\sum_{c}a_ j^c\left[\delta^{ax}\left(\Delta_{ i j}\varepsilon^{xcb}+\frac{\gamma_{ij}}{2}\varepsilon^{ycb}\right)+\delta^{ay}\left(\Delta_{ i j}\varepsilon^{ycb}-\frac{\gamma_{ij}}{2}\varepsilon^{xcb}\right)\right]\nonumber\\
	&+\sum_{c}a_ i^c\left[\delta^{bx}\left(\Delta_{ i j}\varepsilon^{xca}+\frac{\gamma_{ij}}{2}\varepsilon^{yca}\right)+\delta^{by}\left(\Delta_{ i j}\varepsilon^{yca}-\frac{\gamma_{ij}}{2}\varepsilon^{xca}\right)\right]\nonumber\\
	&-\sum_{c}a_ j^b\left[ \left(c_{ i j}^{cx}+a_ i^ca_ j^x\right)\left(\Delta_{ i j}\varepsilon^{xca}+\frac{\gamma_{ij}}{2}\varepsilon^{yca}\right)+\left(c_{ i j}^{cy}+a_ i^ca_ j^y\right)\left(\Delta_{ i j}\varepsilon^{yca}-\frac{\gamma_{ij}}{2}\varepsilon^{xca}\right)\right]\nonumber\\
	&-\sum_{c}a_ i^a\left[\left(c_{ i j}^{xc}+a_ i^xa_ j^c\right)\left(\Delta_{ i j}\varepsilon^{xcb}+\frac{\gamma_{ij}}{2}\varepsilon^{ycb}\right)+\left(c_{ i j}^{yc}+a_ i^ya_ j^c\right) \left(\Delta_{ i j}\varepsilon^{ycb}-\frac{\gamma_{ij}}{2}\varepsilon^{xcb}\right)\right]\nonumber\\
	&+\sum_{c}a_ i^c\sum_{ m\neq i j}\left[\left(c_{ m j}^{xb}\Delta_{ i m}-c_{ m j}^{yb}\frac{\gamma_{ i m}}{2}\right)\varepsilon^{xca}+\left(c_{ m j}^{yb}\Delta_{ i m}+c_{ m j}^{xb}\frac{\gamma_{ i m}}{2}\right)\varepsilon^{yca}\right]\nonumber\\
	&+\sum_{c}a_ j^c\sum_{ m\neq i j}\left[\left(c_{ i m}^{ax}\Delta_{ j m}-c_{ i m}^{ay}\frac{\gamma_{ j m}}{2}\right)\varepsilon^{xcb}+\left(c_{ i m}^{ay}\Delta_{ j m}+c_{ i m}^{ax}\frac{\gamma_{ j m}}{2}\right)\varepsilon^{ycb}\right]\nonumber\\
	&+\sum_{c}\sum_{ m\neq i j}\left[\left(c_{ i j m}^{cbx}\Delta_{ i m}-c_{ i j m}^{cby}\frac{\gamma_{ i m}}{2}\right)\varepsilon^{xca}+\left(c_{ i j m}^{cby}\Delta_{ i m}+c_{ i j m}^{cbx}\frac{\gamma_{ i m}}{2}\right)\varepsilon^{yca}\right]\nonumber\\
	&+\sum_{c}\sum_{ m\neq i j}\left[\left(c_{ i j m}^{acx}\Delta_{ j m}-c_{ i j m}^{acy}\frac{\gamma_{ j m}}{2}\right)\varepsilon^{xcb}+\left(c_{ i j m}^{acy}\Delta_{ j m}+c_{ i j m}^{acy}\frac{\gamma_{ j m}}{2}\right)\varepsilon^{ycb}\right].\nonumber
	\end{align}
	\endgroup
\end{widetext}
These equations form a set of coupled ordinary differential equations, which can be integrated by standard numerical methods.

We calculate the steady-state density operator $\rho_\text{ss}$ by setting to zero the left-hand sides of \eqref{ca_eq1} and \eqref{ca_eq2}, and numerically solving the resulting algebraic equations by a standard Newton-Krylov solver. The stationary values of the $a$- and $c$-coefficients encode the required information on $\rho_\text{ss}$.

\section{Benchmarking against exact results}
\label{appx:benchmarking}

The accuracy of the proposed simulation method is tested by benchmarking the fluorescence spectrum $S$ against exact results. Up to $N=7$ spins (atoms) could be dealt with exactly by using the ``Quantum Toolbox in Python''~\cite{Johansson2012,Johansson2013}, a module tailored to simulate the dynamics of open quantum systems and especially those of quantum optics. 

\begin{figure}
	\centering
	\includegraphics[width=1\linewidth]{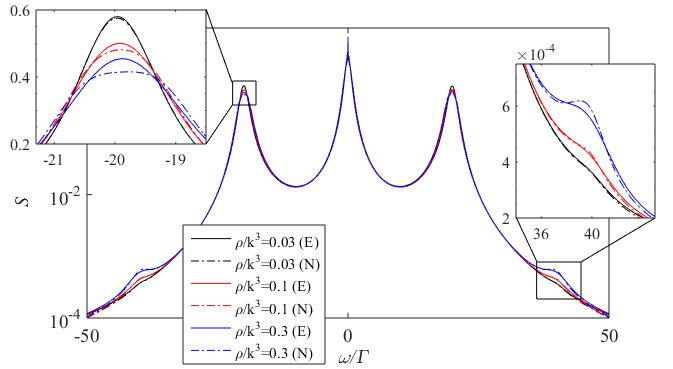}
	\caption{\label{fig:BMO20} Fluorescence spectrum for a cloud of $N=7$ atoms, driven at $\Omega_0=20\Gamma$ at resonance ($\Delta=0$), and for densities $\rho/k^3=0.03$ (black), $0.1$ (red) and $0.3$ (blue). Solid lines refer to exact results (E), dash-dotted lines to the simulation technique described in this Supplemental Material (N). The left inset shows, on a linear scale, a main Mollow sideband, the right inset magnifies one of the novel secondary sidebands.}
\end{figure}

As shown in Fig.~\ref{fig:BMO20} for Rabi frequency $\Omega_0=20\Gamma$, for densities up to $\rho/k^3=0.1$ our simulation results are in very good agreement with exact results, for the main as well as the secondary Mollow sidebands. For larger densities ($\rho/k^3=0.3$ in that same figure), when the coupling gets stronger and higher order correlations are expected to become more relevant, the two spectra exhibit more substantial deviations, although the agreement is still acceptable. Other values of the driving frequency $\Omega_0$ lead to a similar degree of agreement (not shown).

Besides the spectra, we also benchmarked other relevant quantities, including the steady state $\rho_\text{ss}$ calculated according to Sec.~C, as well as the the two-time correlations evolved from the latter. All show very good agreement for densities up to $\rho/k^3\sim0.3$.

\bibliography{refs}

\end{document}